# Control Informed Design of the IAC Autonomous Racecar for Operation at the Dynamic Envelope

Qilun Zhu, Matthias Schmid, Robert Prucka, Ashley Boncimino, and Chris Paredis

*Abstract*— This article introduces the hardware-software co-design of the control system for an autonomy-enabled formula-style high-speed racecar that will be utilized as the deployment platform for high-level autonomy in the first ever head-to-head driverless race called the Indy Autonomous Challenge. The embedded control system needs to facilitate autonomous functionality, including perception, localization, and by-wire actuation, at high speeds and dynamic limits of the vehicle. Rapid maneuvering during the race, however, excites transient dynamics of the vehicle and the actuators. Compared to current autonomous driving focused on highway cruising and urban traffic, transient vehicle control imposes new challenges to the algorithm and system design. The presented work introduces the cascaded control structure employed by the IAC prototype to fully exploit the time scale separation between different control tasks. It is demonstrated by example way how the model-based control strategies and simulation are utilized to inform the decisions in the actuation, computation, perception, and software pipeline design decisions for the first-of-its-kind IAC racecar.

## I. INTRODUCTION

The rapid development of Autonomous Driving (AD) technology promises to change the landscape of transportation. Competitions involving driving situations near the dynamic envelope have been proven to be a catalyst for the maturation of technologies. Since the DARPA Grand Challenge [1], many competition events of various scales were hosted to boost the development of autonomous driving technology, including the DARPA urban challenge [2], Formula Student Driverless [3], and Roborace [4]. The Indy Autonomous Challenge (IAC) seeks to motivate AD technology development through the first ever head-to-head high-speed autonomous race. University teams worldwide will deploy their developed algorithms on formula-style vehicles – equipped with the identical hardware design – and compete on the oval track of the Indianapolis Motor Speedway (IMS). To win the challenge, a team needs to outrun, outmaneuver, and outsmart their competitors, requiring highly complex computations completed within milliseconds. These extreme or 'edge-case' driving scenarios offer a rare opportunity to prove that AD can control vehicles at least as safe as human drivers. With the promises of increased safety (here in the sense of stably handling the vehicle in extreme situations) and energy efficiency, advocating AD technologies by linking them to competition and motorsports will help to raise acceptance in the court of public opinion. This paper introduces the control system design of the IAC racecar and the critical role of model predictive control in the decision-making process.

The vehicular platform, i.e., the high-speed, autonomy-enabled, open-wheel racecar for the IAC is developed and prototyped through the Deep Orange 12 project at Clemson University based on a Dallara IL-15 chassis. This allows for the teams participating in the IAC to solely focus on the development of their AD software, which will then be executed on the dedicated Competition Team Computer in the Clemson-designed vehicle. Built on a formula-style carbon fiber monocoque chassis with an open cockpit and a racing setup for the oval track at IMS, this vehicle achieves a power-to-weight ratio of 0.5 kW/kg and a top speed of 290 km/h. During cornering and overtaking maneuvers, the car can change its heading direction (or yaw angle) at a rate of 45 deg/sec while coping with approximately 2.8 Gs of lateral acceleration. The design process for this vehicle involved considerations for a large variety of topics: structural calculations, powertrain selection and modification, electrical system design, sensor selection and placement, emergency procedures, communications links, robust navigation, formulation of race rules, etc. Within the limited scope of this article, we only focus on the hardware-software co-design by illustrating how control considerations informed design decisions.

The dynamic envelope of a vehicle is limited by its mechanical capacities, including tire traction, engine power, etc. During highly dynamic vehicle operations such as obstacle avoidance and overtaking, the envelope shrinks due to transient motions. For example, in a fast lane change scenario during which the vehicle needs to alter its heading direction twice, the rear tires might initiate skidding during the second turn due to unsettled body roll and yaw motion. To ensure safe vehicle operation, the control system must be capable of identifying obstacles at sufficient distances and of quickly computing a feasible trajectory considering the vehicle's transient dynamic envelope. Previous research, which is also utilized by some teams competing in the IAC, employs receding horizon Model Predictive Control (MPC) for vehicle motion planning (e.g. [5], [6] and [7]). Motivated by this research, we utilize modeling and simulation at high speed to determine the car's maneuver capability under an MPC scheme. This informs us on the limits of perception, localization, and planning necessary to execute drastic maneuvers. These results are then incorporated into the requirements for the control system and instrumentation hardware and software design enabling a successful IAC for the teams. This paper does not address the particular perception and decision-making processes, as their development is the core challenge of the IAC. The dynamics

All authors are with the Department of Automotive Engineering, Clemson University, Greenville, SC 29607. e-mail: Qilun Zhu: qilun@clemson.edu; Matthias Schmid: schmidm@clemson.edu; Ashley Boncimino: aboncim@clemson.edu; Robert Prucka: rprucka@clemson.edu; Chris Paredis: paredis@clemson.edu

and delay of actuators significantly affect vehicle behavior at high speed. Due to the slow update frequency of the vehicle control, these effects are often lumped together as unobservable disturbances and handled by robust control strategies. Robust MPC (RMPC) has been developed to solve constrained optimal control problems under the influence of uncertainties. This control strategy has been widely adopted for flight control of UAVs and airplanes, [9], [10], and [11]. The optimization problem solving for the desired control actions then can be formulated as a min-max problem. Although the RMPC strategy was proven to be closed-loop stable, solving the potentially min-max problem in real-time is often impractical. Several methodologies were developed to simplify the min-max problem to determine suboptimal solutions for real-time control applications. Tube-based MPC (TMPC) is such an example[12]. It can achieve the additional robustness requirement by solving a simplified optimal control problem with the same order of complexity as conventional MPC. This research applies TMPC to handle the uncertain response of the steer-by-wire system and quantifies its impact on the minimum perception distance. Please note that the vehicle models are based on the combination of analysis by the Deep Orange 12 team and data from the contributing sponsors. Due to proprietary reasons, the models, data, and parameters cannot be shared outside the IAC at this point.

## II. CASCADED VEHICLE CONTROL SYSTEM DESIGN

The vehicle's control, computational, sensory, and communication system needs to provide necessary capabilities for all autonomous driving tasks, including environment perception, localization, trajectory planning, decision making, and control execution. While most autonomous driving vehicles enable these features to a certain degree, the high vehicle speed in a competition against other cars imposes significant design challenges for the computation and execution time of the perception-control pipeline as well as for its execution accuracy. The participating cars must follow race control commands through wireless communication and handle emergency situations such as accidents and hardware malfunctions. The underlying control tasks have different levels of criticality for safe vehicle operation. They also must cope with subsystems exhibiting dramatically different time scales, e.g., from camera trigger synchronization with microsecond precision to servo motor dynamics in the millisecond range to tire temperatures varying within tens of seconds. A cascaded control structure is therefore employed to properly allocate computational resources to these control tasks. The hardware architecture of the sense-think-act pipeline consists of several computational platforms and is shown in Figure 1. The 'intelligence' capabilities, i.e., most of the higher-level autonomous driving features, are realized on a dedicated Autonomy Computer (AC), also called Competition Team Computer, which is the only hardware component that is accessible to the participating teams. The software deployed on this platform will be the sole responsibility of the competition teams to close the sense-think-act loop. The connection to the dynamics layer must be realized via a mandatory ROS-CAN node that defines the vehicle interface with standardized topics. To minimize latency, to maximize data transfer, and to enable full flexibility in sensor configuration, all exteroceptive sensors are connected to the AC either directly via CAN or via an Ethernet connection. The Ethernet traffic is handled by a switch that also provides time synchronization for the connected devices (Lidars, cameras, AC, GNSS, Wireless Ethernet).

The dynamics layer software architecture communicates with the AC through the CAN protocol. For braking and steering demands, the dynamics controller tracks the actuation demand from the AC by regulating the servo motor position setpoints. The front and rear braking pressure demands are first translated into brake master cylinder travel with the subsequent motor position through lookup tables that have been calibrated under a nominal condition. Braking pressure feedback is then utilized to correct for pressure variations and to compensate for delays. A similarly structured feedback control of the steering angle is needed to eliminate steady-state offset and to improve motor response. The powertrain demand passes through the dynamics layer to the Engine Control Unit (ECU), which coordinates the engine and transmission actuators.

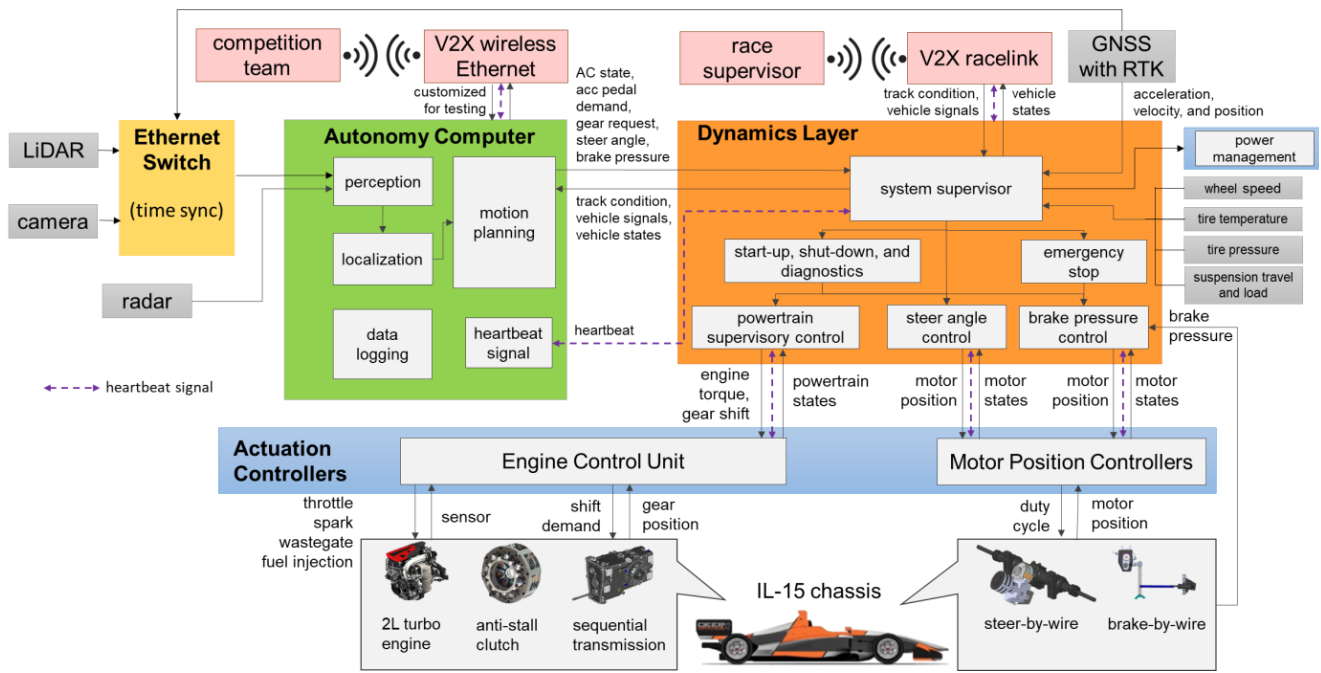

**Figure 1.** Control system architecture of the autonomy-enabled Deep Orange 12 prototype

On another level, the dynamics layer controller also provides vehicle diagnostics and state estimation, as it can sample sensors related to vehicle dynamics at a faster rate than the AC. This information includes vehicle wheel speed, tire temperature and pressure, suspension travel and load, braking pressure, and steering angle. A secondary (CAN) feed of the Global Navigation Satellite System (GNSS) with Real-Time Kinematic (RTK) and incorporated IMU is connected to the dynamics controller (in addition to the Ethernet connection with the AC) to inform on the vehicle's location, speed, and acceleration at 100 Hz. The dynamics layer has access to powertrain sensors through CAN communication with the ECU. Whereas the dynamics controller reports most sensor information to the competition team computer (accessible via wireless Ethernet), the dynamics controller's faster sampling rate allows for closer monitoring of mechanical and software failures. In extreme cases of these failures, the dynamics layer bypasses the autonomy computer and executes an emergency stop procedure. The controller also observes the race control communication via the Mylaps Racelink system. While the competition teams are primarily responsible for executing emergency maneuvers (including coordinated emergency stops and evasive maneuvers), the dynamics layer is responsible for a secondary rudimentary emergency stop and shut-down procedure (called Purple Flag) that can either be triggered by the AC, by race control, or by loss of heartbeat signal to the AC (automatic trigger). As the tasks handled by the dynamics layer are vital to safe vehicle operation, it is required to execute codes with high reliability and accurate timing. On the physical side, the controller also needs to be robust against vibration, humidity, and electromagnetic interference. As a result, the selected control deployment pipeline optimizes the program and pre-compiles it into machine code with clearly defined execution priority on a real-time control platform. The control system has three CPU cores running at 260MHz and 4 MB of flash space, packaged in a sealed aluminum casing with thermal and vibration insulation. The actuation layer of the DO12 control architecture consists of an ECU and two separate motor position controllers for the by-wire systems. The controllers in the actuation layer track the setpoints with even faster update frequencies than the dynamics layer. Here, PID controllers are employed to track desired positions for the steering and braking servo motors utilizing the feedback from angular potentiometers. The sensors are instrumented on the output shaft of the gearbox which reduces the servo's travel to the desired magnitude (less than one rotation for both braking and steering). The turbocharged engine requires feedback control to ensure stable operation and to reduce turbo lag. This management involves coordinating throttle, wastegate, fuel injection, and spark timing to deliver the desired engine torque output while preventing compressor surge, misfire, and engine knock. The selected ECU also controls the gear shift of the employed sequential type transmission. The necessary engine torque management during the gear shift period is pre-programmed in the ECU and includes a launch control functionality to engage the centrifugal clutch when starting the vehicle from rest.

### III. APPLICATIONS OF MODEL PREDICTIVE VEHICLE CONTROL IN DESIGN DECISIONS

#### A. Minimum control update frequency

The MPC-based vehicle control relies on the availability of a control-oriented dynamics model to predict the vehicle's behavior for a certain period into the future. The suspension of the IL-15 vehicle is tuned to minimize body motion and variation of aerodynamic forces. Therefore, roll and pitch motions are neglected for the analysis. However, the transfer of tire vertical load caused by roll and pitch momentum is captured by the model. The equations of motion in the vehicle body frame are:

$$m(\dot{u} - \dot{\psi}v) = \sum F_x \quad (1)$$
$$m(\dot{v} + \dot{\psi}u) = \sum F_y \quad (2)$$
$$I_{zz}\ddot{\psi} = \sum M_z \quad (3)$$
$$\dot{X} = u\cos\psi - v\sin\psi \quad (4)$$
$$\dot{Y} = u\sin\psi + v\cos\psi \quad (5)$$

where $m$ is the mass of the vehicle; $u$ is the longitudinal velocity; $\psi$ is the yaw angle; $v$ is the lateral velocity; $F_x$ is external force exerted on the CG along the longitudinal direction in the body frame; $F_y$ is external force exerted on the CG along the lateral direction in the body frame; $I_{zz}$ is the rotational inertia of the z-axis; $M_z$ is the moment of external force forces exerted on the CG of the vehicle in the body

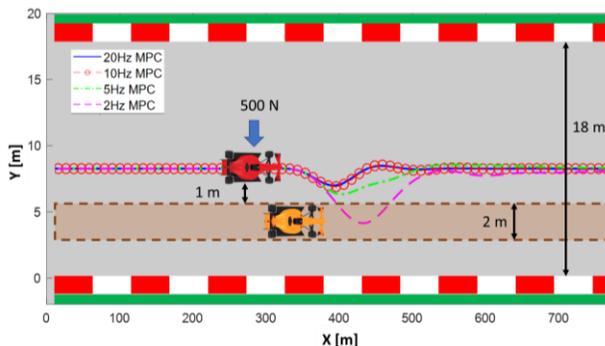

**Figure 2.** Comparison between MPCs for an external disturbance with different update frequencies in the scenario of overtaking

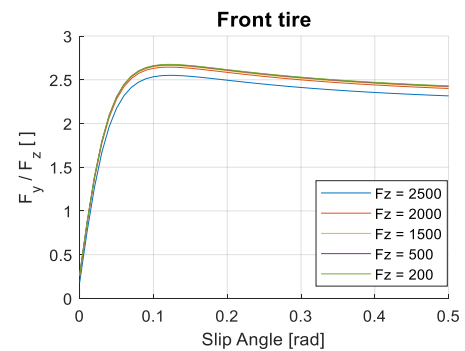

**Figure 3.** Nondimensional tire friction versus slip angle for front axle based on provided manufacturer data

frame; $X$ and $Y$ are the coordinates of the vehicle in the inertial frame.

The tire forces in the vehicle model are computed using the Pacejka model provided by the supplier. At racing speed, the wheel steering angle and slip angles of tires and body are often smaller than 5 degrees, allowing for a small angle approximation of the vehicle model. Assuming the longitudinal vehicle velocity is maintained by the autonomous driving control, the vehicle dynamics can be linearized at given nominal conditions as a second-order state-space model:

$$\dot{x} = Ax + B\delta \quad (6)$$

where $x = [\psi, \dot{\psi}, \beta, Y]^T$; $\beta = \tan^{-1}\frac{v}{u}$; $\delta$ is the steering column angle; $A \in \mathbb{R}^{4\times 4}$ and $B \in \mathbb{R}^{4\times 1}$.

The vehicle has an oversteering tendency at 180 mph, which is the designed top speed for the IAC. This speed is below the critical speed by a significant margin that can be derived from Eq. (6). Discretizing Eq. (6) at this speed with the Zero Order Hold method yields:

$$x^+ = A_d x + B_d \delta \quad (7)$$

where $A_d = e^{A\Delta t}$ and $B_d = \int_0^{\Delta t} e^{A\tau} d\tau \, B$; $\Delta t$ is the sampling period of the vehicle control.

The calculation of $B_d$ becomes numerically intractable for $\Delta t > 0.6s$, which corresponds to a minimum update frequency of 1.67 Hz. For a control frequency higher than this threshold, the vehicle trajectory and lateral motion are controllable through steering. However, feasible control solutions may not exist due to constraints on the trajectory and tire friction. To investigate the minimum achievable control update frequency, we consider an overtaking scenario as shown in Figure 2. In this scenario, the fast-moving vehicle passes the slower vehicle, which is holding the inside lane. The control update frequency must allow the MPC-controlled vehicle to maintain its overtaking path under the influence of external disturbances without crashing into the other car. The MPC problem solving for the desired steer action is formulated as:

$$\min_{\delta_k, \delta_{k+1}, \dots, \delta_{k+N-1}} x_{k+N}^T Q_N x_{k+N} + \sum_{i=k+1}^{k+N-1} (y_i - y_{ref})^T Q_y (y_i - y_{ref}) + \delta_i^T Q_u \delta_i \quad (8)$$

subject to: $x^+ = A_d x + B_d \delta$
$Y_{min} \leq Y_j \leq Y_{max}$
$\left|\beta_j + \frac{\dot{\psi}_j}{u}a - \delta_j\right| < \alpha_{max}$
$\left|\beta_j - \frac{\dot{\psi}_j}{u}b\right| < \alpha_{max}$
$|\delta_j| < \delta_{max}$

where $k$ is the index of the current time step; $Q_N$, $Q_y$, and $Q_u$ are the weights on terminal state, output deviation, and control effort; $N$ is the number of time steps of the preview horizon; $y = [Y, \psi]^T$; $j = k+1, k+2, \dots, k+N$; $Y_{min}$ and $Y_{max}$ are the minimum and maximum of the vehicle's Y coordinate; $a$ and $b$ are the CG's distance to the front and rear axles; $\alpha_{max}$ is the maximum slip angle of the tires; $\delta_{max}$ is the maximum steering angle.

The tire friction limit is imposed indirectly by limiting the slip angle of both axles. Figure 3 shows the relationship between tire friction coefficient versus slip angle for both axles. It can be observed that the friction coefficient is not sensitive to the vertical load of the tire. Constraining the axle slip angles can prevent the vehicle from sliding. The MPC problem of Eq. (8) is solved using an active set Quadratic Programming algorithm for every control update [9]. Figure 2 depicts the simulation result of the investigated overtake scenario, generated using the model described by Eq. (1)-(5) and Matlab Simulink. An external lateral force of 500 N is applied to the overtaking vehicle for 0.5 seconds, thus mimicking a disturbance such as in a crosswind situation that causes the vehicle to deviate from its trajectory. The reference trajectory in this scenario is a straight line with a constant distance to the side of the race track (based on the assumption that the raceline is known). The simulation compares the resulting MPC-based vehicle control for four different update frequencies. The four MPC laws share the same preview distance and penalty terms in the cost function. It can be observed that all four MPCs can stabilize the vehicle under the applied disturbance. For frequencies less than 10 Hz, however, the MPC cannot determine a trajectory that does not intersect with the trajectory of the slow-moving vehicle. These results can now be utilized to directly inform the hardware-software co-design: the actual update frequency of the IAC vehicle depends on the computational efficiency of the perception and decision-making program coded by the competition teams. It is well documented that finishing these computations in 50 – 100 ms is very challenging and requires state-of-the-art high performance computing. The dynamics layer and actuation controllers are designed to run at 100 Hz (for both control and measurement purposes) to provide sufficient frequency separation with the vehicle control. This separation allows for signal processing that improves measurement accuracy. It also reduces communication delays and the resulting uncertainties.

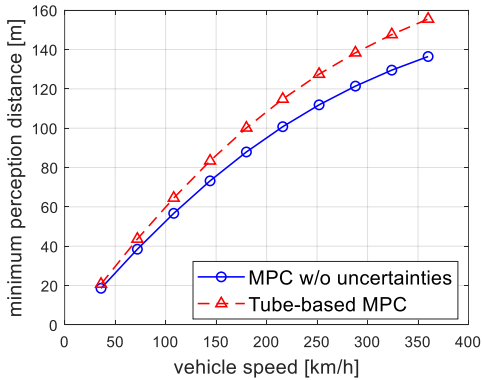

**Figure 4.** Minimum safe perception distance vs. vehicle speed

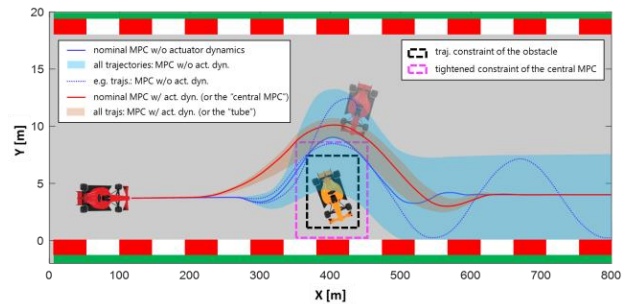

**Figure 5.** Comparison between different MPC based vehicle control strategies under the obstacle avoidance simulation

### B. Minimum perception distance

The MPC problem described by Eq. (8) is also used to investigate the necessary minimum perception distance to avoid an obstacle. The obstacle is incorporated in the MPC in terms of constraints on the vehicle's coordinates. Figure 5 illustrates the simulation results of the autonomous racecar attempting to avoid a broken-down vehicle at a speed of 290 km/h. The box with a black dashed outline depicts the trajectory constraint imposed by the blocking vehicle. The solid blue line corresponds to the trajectory generated by the MPC with a 1.5-second preview horizon, which can be converted into a perception distance of 121 meters at the given speed. A preview horizon shorter than this threshold will render the MPC problem infeasible. Repeated investigation for different vehicle speeds leads to the non-surprising observation in Figure 4 that minimum perception distance increases with vehicle speed. The perception sensors selected for a given vehicle must therefore enable the minimum detection distance requirement occurring at the highest speed of autonomous operation. The actuator dynamics and communication/computation delays significantly reduce the phase and gain margin of the closed-loop vehicle control in a highly transient operation such as the obstacle avoidance scenario discussed above. Neglecting these delays can lead to unstable vehicle behavior. The blue shaded area in Figure 5 comprises all possible vehicle trajectories generated by simulating the MPC strategies of Eq. (8) with a high-fidelity vehicle model (Eq. (1)-(5) with added steering dynamics) for over 1000 repetitions. During each repetition, a random realization of the actuator delay is imposed on the vehicle model via a Monte Caro approach. In comparison to the simulation with the ideal vehicle model (blue solid line in Figure 5), the added actuator dynamics and delay in the simulation cause the vehicle to deviate significantly from its nominal trajectory. Consequently, the vehicle could potentially hit the obstacle with a probability of approximately 29%. Furthermore, the dotted blue lines demonstrate two other extreme cases out of all possible trajectories for the conventional MPC. Here, the vehicle spins out of control in the first case when the controller attempts to return the car back to the optimal race line. In the second case, the MPC is unable to stabilize the vehicle after avoiding the obstacle. Further investigation reveals that the instability for the latter case is caused by a phase lag induced by the steering actuation.

### C. Tube-based MPC

The steer-by-wire systems must overcome tire friction and alignment torque in addition to the corresponding inertia to provide the required actuation. Figure 6 details an example of steering actuation response while tracking a step decrease in steering angle demand during a bench test with the deployed hardware. The transport delay is caused by the communication between different control levels and the time required to compute the control action. Further, more extensive tests have revealed that the communication delay between the AC and steer actuation (two layers of CAN communication at 100 Hz) ranges from 10 to 25 milliseconds. The computation time of the proposed MPC ranges from 5 to 20 milliseconds. The actual autonomous driving control developed by the competition teams will likely take more time due to the added perception function. Our analysis assumes that the computation can be finished within 100 milliseconds based on the minimum frequency discussion, resulting in a range of total transport delay from 15 to 125 milliseconds. The subsequent delay of the first-order appearance is the result of the coupled motor dynamics and position feedback control. The time constant of this delay varies with the steering load, ranging from 50 to 500 milliseconds.

Even with the high-power computation platform available on our autonomous racecar, it is challenging to increase the control frequency beyond 20 Hz. Yet, this sampling frequency is too low to capture the details of steering dynamics. In practical terms, the steering dynamics are reduced to simple transfer functions with input transport delay in this analysis. Errors of the nominal actuator dynamics and delays can therefore be treated as parametric uncertainties with upper and lower bounds. The following equations show the linearized and discretized system dynamics with augmented steering state and disturbance:

$$\hat{x}^+ = \hat{A}\hat{x} + \hat{B}\delta \tag{9}$$

where $\hat{x} = [\psi, \dot{\psi}, \beta, Y, x_\delta, x_{delay}]^T$; $\hat{A} \in \mathbb{R}^{6\times 6}$ and $B \in \mathbb{R}^{6\times 1}$ are from linearizing the system dynamics augmented with steering dynamics and delay; parameter $p \coloneqq (\hat{A}, \hat{B})$ can, at any time, take any value in the convex set $\mathcal{P}$ defined by $\mathcal{P} \coloneqq co\{(\hat{A}_j, \hat{B}_j) | j \in \mathcal{J}\}, \mathcal{J} \coloneqq \{1,2,...,J\}$. $\mathcal{P}$ is designed to represent the range of steering response time and delay.

Given the focus on vehicle design and the limited scope of this paper, we introduce the TMPC formulation only briefly. Details of the applied TMPC method and its stability proof can be found in [14]. The TMPC splits the control strategy into a central MPC and a local feedback law. The optimal control of the central MPC can be solved via:

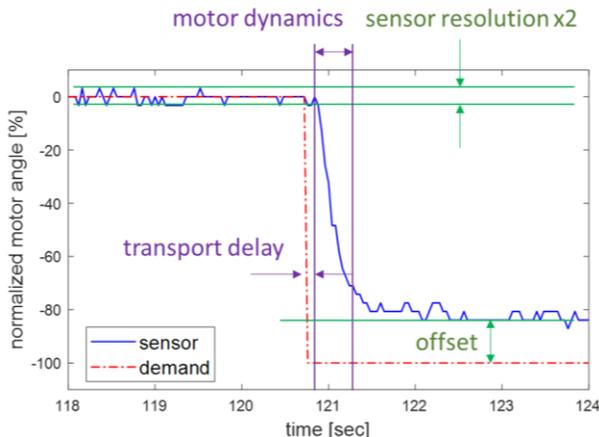

**Figure 6.** An example of steer-by-wire response tracking a step change of steering angle demand during a bench test

$$\Sigma^* = \underset{\sigma_k, \sigma_{k+1}, \ldots \sigma_{k+N-1}}{\operatorname{argmin}} z_{k+N}^T Q_N z_{k+N} +$$

$$\sum_{i=k+1}^{k+N-1} (y_i - y_{ref})^T Q_y (y_i - y_{ref}) + \sigma_i^T Q_u \sigma_i \quad (10)$$

$$\text{subject to:} \quad z^+ = \bar{A}z + \bar{B}\sigma$$
$$z \in \mathbb{Z} \text{ and } \sigma \in \mathbb{V}$$

where $\Sigma = [\sigma_k, \sigma_{k+1}, \ldots \sigma_{k+N-1}]^T$; $\bar{A} = \frac{1}{J}\sum_{j=1}^J A_j$ and $\bar{B} = \frac{1}{J}\sum_{j=1}^J B_j$; $z \in \mathbb{R}^{6\times 1}$ and $\sigma \in \mathbb{R}^{1\times 1}$ are the nominal state and control used in the central MPC; $\mathbb{Z} \coloneqq \mathbb{X} \ominus S_K(\infty)$ and $\mathbb{V} \coloneqq \mathbb{D} \ominus S_K(\infty)$; $\mathbb{X}$ and $\mathbb{D}$ are the feasible set of states and control defined in Eq. (8); $S_K(\infty)$ is the invariant uncertainty set determined by the local feedback law; $\ominus$ represents set subtraction.

The invariant uncertainty set $S_K(\infty)$ that is used to tighten the state and control constraints of the central MPC is computed numerically by:

$$S_K(\infty) = \lim_{i \to \infty} S_K(i) \quad (11)$$

where $S_K(i) \coloneqq \sum_{j=0}^{i-1} A_K^j \mathbb{W}$; $A_K \coloneqq \bar{A} + \bar{B}K$; $K$ is the linear state feedback control gain obtained through pole placement; $\mathbb{W} \coloneqq \{(\hat{A} - \bar{A})x + (\hat{B} - \bar{B})u \mid (\hat{A}, \hat{B}) \in \mathcal{P}, (x, \delta) \in \mathbb{X} \times \mathbb{D}\}$.

The local state feedback control law is linear and designed through pole placement, manually tuned using the Nyquist method (stable for different parameters). For our selected $K$, $S_K(i)$ stops changing after $i > 7$. Finally, the complete control law can be expressed as:

$$\delta^* = \sigma^* + K(x - z) \quad (12)$$

where $\sigma^*$ is the first element of $\Sigma^*$.

The solid red line in Figure 5 shows the nominal vehicle trajectory generated by the central MPC. When the vehicle deviates from the nominal model due to disturbances, the local feedback controller then steers the vehicle back to the nominal trajectory. For all possible disturbance traces, the proposed vehicle control results in a 'tube' of trajectories around the 'center', illustrated by the red area in Figure 5. Here, the local feedback control restricts the 'size' of the tube by $S_K(\infty)$. For vehicle control, one of the tightened constraints can be visualized similarly to a safety margin surrounding the incoming obstacle, as it is shown via the magenta box in Figure 5. The tightened constraints 'enlarge' the obstacle by approximately 1.5 meters, causing the MPC to react earlier, thus arriving at a feasible trajectory. For the simulated scenario, the central MPC requires a perception distance of 140 meters to remain clear of the enlarged obstacle, thus necessitating 19 meters more than the ideal MPC solution neglecting steering dynamics and delays. The resulting sensitivity of the minimum safe perception distance under TMPC with respect to vehicle speed is depicted as the red line in Figure 4, clearly showing the increased demand for perception distance when considering steer-by-wire responses.

E.  CONCLUSIONS

The Deep Orange team has been facing a hardware-software co-design challenge for the prototype in a high-speed driverless head-to-head racing event that is the first of its kind. Experience and existing data for such a design are therefore strictly limited. This article has provided an overview on the resulting architecture with an emphasis on the cascaded control architecture. We have demonstrated by example how modeling, simulation, and control have been used simultaneously to inform control system design and instrumentation choices for an open-wheel racecar to compete in the IAC. It is the goal of this design to enable future autonomous functionality to be carried out at the dynamic limit on a dedicated autonomy computer, fully accessible to the competition teams for programming. Here, model-based vehicle control played a critical role in providing deep physical insights. Due to confidentiality agreements, code is not openly shared at this point, but we hope that it can be made available in the future.